\documentclass[12pt]{article}

\usepackage{pb-diagram}
\usepackage{latexsym}
\usepackage{amsfonts}
\usepackage[usenames,dvipsnames]{xcolor}
\usepackage{graphicx,amssymb,amsmath,epsfig}
\usepackage[english]{babel}
\usepackage{graphicx}
\usepackage{dcolumn}
\usepackage{bm}

\definecolor{myblue}{rgb}{0,0,0.8}
\usepackage[colorlinks=true
,urlcolor=myblue	
,anchorcolor=myblue
,citecolor=myblue
,filecolor=myblue
,linkcolor=myblue
,menucolor=myblue
,pagecolor=myblue
,linktocpage=true  
]{hyperref}

\catcode`\@=11
\def\marginnote#1{}

\newcount\hour
\newcount\minute
\newtoks\amorpm
\hour=\time\divide\hour by60 \minute=\time{\multiply\hour by60
\global\advance\minute by-\hour}\edef\standardtime{{\ifnum\hour<12
\global\amorpm={am}%
        \else\global\amorpm={pm}\advance\hour by-12 \fi
        \ifnum\hour=0 \hour=12 \fi
        \number\hour:\ifnum\minute<10
0\fi\number\minute\the\amorpm}}
\edef\militarytime{\number\hour:\ifnum\minute<10 0\fi\number\minute}

\def\draftlabel#1{{\@bsphack\if@filesw {\let\thepage\relax
   \xdef\@gtempa{\write\@auxout{\string
      \newlabel{#1}{{\@currentlabel}{\thepage}}}}}\@gtempa
   \if@nobreak \ifvmode\nobreak\fi\fi\fi\@esphack}
        \gdef\@eqnlabel{#1}}
\def\@eqnlabel{}
\def\@vacuum{}
\def\draftmarginnote#1{\marginpar{\raggedright\scriptsize\tt#1}}
\def\draft{\oddsidemargin -.5truein
        \def\@oddfoot{\sl preliminary draft \hfil
        \rm\thepage\hfil\sl\today\quad\militarytime}
        \let\@evenfoot\@oddfoot \overfullrule 3pt
        \let\label=\draftlabel
        \let\marginnote=\draftmarginnote

\def\@eqnnum{(\theequation)\rlap{\kern\marginparsep\tt\@eqnlabel}%
\global\let\@eqnlabel\@vacuum}  }

\def\numberbysection{\@addtoreset{equation}{section}
        \def\theequation{\thesection.\arabic{equation}}}

\def\underline#1{\relax\ifmmode\@@underline#1\else
 $\@@underline{\hbox{#1}}$\relax\fi}

\catcode`@=12 \relax

\numberbysection

\topmargin by -\headsep
\def\nonu{\nonumber}
\def\br{\begin{eqnarray}}
\def\er{\end{eqnarray}}

\def\({\left(}
\def\){\right)}
\def\[{\left[}
\def\]{\right]}

\def\arctanh{\mathrm{Arctanh}}
\def\a{\alpha}

\def\b{\beta}

\def\d{\delta}

\def\bpsi{\bar{\psi}}

\def\eps{\epsilon}

\def\G{\Gamma}

\def\k{\kappa}
\def\l{\lambda}
\def\L{\Lambda}
\def\m{\mu}

\def\o{\over}
\def\om{\omega}

\def\O{\Omega}
\def\p{\phi}
\def\P{\Phi}
\def\pa{\partial}

\def\s{\sigma}
\def\S{\Sigma}
\def\sech{\mathrm{sech}}
\def\t{\tau}
\def\th{\theta}

\def\tp0{\Theta_{+}^{(0)}}
\def\tm0{\Theta_{-}^{(0)}}

\def\z{\zeta}

\def\vp{\varphi}

\def\vep{\varepsilon}
\def\bvep{\bar{\varepsilon}}

\begin{document}

\vspace*{1cm}
\noindent

\vskip 1 cm
\begin{center}
{\Large\bf  ${\cal N}=1$ super sinh-Gordon model in the half line: Breather solutions}
\end{center}
\normalsize
\vskip 1cm
\begin{center}
{A.R. Aguirre}\footnote{\href{mailto:aleroagu@ift.unesp.br}{aleroagu@ift.unesp.br}}{$^\clubsuit$}, J.F. Gomes\footnote{\href{mailto:jfg@ift.unesp.br}{jfg@ift.unesp.br}}{$^\clubsuit$},  L.H. Ymai\footnote{\href{mailto:leandro.ymai@unipampa.edu.br}{leandro.ymai@unipampa.edu.br}}{$^\spadesuit$}and A.H. Zimerman\footnote{\href{mailto:zimerman@ift.unesp.br}{zimerman@ift.unesp.br}}{$^\clubsuit$}\\[.7cm]

\par \vskip .1in \noindent
{$^\clubsuit$} {Instituto de F\'isica Te\'orica - IFT/UNESP,\\
Rua Doutor Bento Teobaldo Ferraz, 271, Bloco II,
CEP 01140-070,\\ S\~ao Paulo-SP, Brasil.}\\[0.3cm]
{$^\spadesuit$} {Universidade Federal do Pampa - UNIPAMPA,\\
Travessa 45, 1650, Bairro Malafaia, CEP 96413-170, Bag\'e-RS, Brasil.}
\vskip 2cm

\end{center}

\begin{abstract}

We examine the ${\cal N}=1$ super sinh-Gordon (SShG) model restricted into the half line through a reduction from the defect SShG model. The B\"acklund transformations are employed to generate one-, two- and three-soliton solutions as well as  a class of breathers solution for this model. The parameters of such classical solutions are shown to satisfy some contraints in order to preserve both integrability and supersymmetry properties of the original bulk theory. Additionally,  previous results are recovered when performing the purely bosonic limit.

\end{abstract}

\newpage
\tableofcontents

\vskip .4in

\section{Introduction}

The study of nonlinear classical integrable field theories  with boundaries  has started with the work of Sklyanin \cite{Sklyanin1,Sklyanin2}, Tarasov \cite{Tarasov}, and  Habibullin \cite{Khabi, Khabi2, Khabi3} within the context of the sine-Gordon model.  
On the other hand, field theories in the half line may be formulated by imposing appropriated integrable boundary conditions, for instance,  at the origin $x=0$.  In \cite{Tarasov, Khabi,Khabi2,Khabi3} the B\"acklund transformation arises as an important tool to consider the structure of the model.  
 In the Lagrangian framework, this is equivalent to impose a suitable boundary potential preserving integrability.
The simplest example is the boundary sinh (sine)-Gordon model \cite{Zamo}, whose Lagrangian density can be written as
\begin{eqnarray}
 {\mathcal L} &=& \theta(-x)\left[\frac{1}{2}(\partial_x \phi)^2 - \frac{1}{2}(\partial_t \phi)^2 + 4m^2 \cosh(2\phi) \right] +\delta (x) \Big[ \Lambda \cosh\left(\phi -\phi_0\right)\Big]. \label{e1.1}
\end{eqnarray}

In \cite{Zamo} it was noticed that, in this case,  the most general boundary condition preserving integrability has two free para\-me\-ters $\L$ and $\phi_0$, namely, 
\begin{equation}
 \partial_x \phi \big|_{x=0}= -\Lambda \sinh\big(\phi -\phi_0\big)\Big|_{x=0}.\label{e1.2}
\end{equation}

Bowcock et al. \cite{bow} proposed a formulation in terms of Lax operators  in the half line such that the boundary conditions derived in \cite{Zamo}  would be reproduced.

In \cite{SSW} Saleur, Skorik and Warner discussed the classical  approach to the boundary sine-Gordon model by extending the method of images  proposed by Cardy \cite{Cardy} and by making use of B\"acklund transformations in order to obtain the boundary conditions derived by Ghoshal and Zamolodchikov \cite{Zamo}, as well as solutions to the problem in the half line.

The supersymmetric extension of the sinh-Gordon model restricted on the half-line has also been studied from the classical point of view of Inami {et al.} \cite{Ina1} and Nepomechie \cite{Nepo1}. In \cite{Ina1} it was claimed that the combined constraints of integrability and supersymmetry do not allow any free parameters in the boundary potential. Nevertheless, some years later it was pointed out in \cite{Nepo1} that the boundary super sinh-Gordon model  actually has a two-parameter family of integrable boundary potentials taking into account  the introduction of  fermionic degrees of freedom. 

More recently Bowcock et al. \cite{Bow}  derived the B\"acklund transformation   through the  study of defects in the bulk for the bosonic  sine-Gordon and  other models within the Lagrangian/Hamiltonian  formulation.  That ensures the integrability and henceforth modified conservation laws  of the systems.   The  sine-Gordon model with  B\"acklund defect at the origin is given by the  Lagrangian density,
\begin{eqnarray}
 {\mathcal L} &=& \theta(-x)\left[\frac{1}{2}(\partial_x \phi_1)^2 - \frac{1}{2}(\partial_t \phi_1)^2 +V_1(\phi_1) \right] + \theta(x) \left[\frac{1}{2}(\partial_x \phi_2)^2 - \frac{1}{2}(\partial_t \phi_2)^2 +V_2(\phi_2) \right] \nonumber \\
 &\mbox{}&+\delta(x)\left[ \frac{1}{2}(\phi_2\partial_t\phi_1-\phi_1\partial_t\phi_2) + 2m\left(\s \cosh(\phi_1+\phi_2) +\frac{1}{\s}\cosh(\phi_1-\phi_2) \right)\right],\label{e1.3}
\end{eqnarray}
where $V_p(\phi_p) = 4m^2\cosh(2\phi_p)$ with $p=1$ if $x<0$ and $p=2$ if $x>0$. In fact, these defect conditions turn out to be ``frozen" B\"acklund transformations for the sinh-Gordon model. It is worth pointing out that alternative approaches have been provided in \cite{Ana1,Ana2,Kundu} to prove Liouville integrability for the defect sine-Gordon model which are based on the classical $r$-matrix language.

The extension of the B\"acklund formulation of these results  to the supersymmetric sine-Gordon and Grassmanian  models were already considered in \cite{Lean1,Lean2,Ale,Ale2,Ale3,Ale4}. Within the B\"acklund transformation  context it appears natural to consider boundaries describing the half line by adapting the  considerations in \cite{Sklyanin1, Tarasov, Khabi,Khabi2,Khabi3, SSW}.

In this paper, we formulate a free fermionic field theory and  the ${\cal N}=1$ supersymmetric sinh-Gordon (SShG) model in the half line by extending the method of images and B\"acklund transformations. We construct explicit soliton and breather solutions for ${\cal N}=1$ SShG. Also, we extend to the supersymmetric case the results obtained in \cite{CorrDelius}, and consider the corresponding supersymmetric breather solutions. By taking a three-soliton solution with one static, we construct another class of supersymmetric breather solution, generalizing therefore the results of Saleur et al. \cite{SSW}.

This paper is organized as follows. In section 2 we review the defect ${\cal N}=1$ SShG model and present the  supersymmetry transformations that leave the action invariant.  In section 3 we discuss the half-line limiting procedure for the bosonic sine-Gordon and free fermionic theory.  In this last case  we reproduce the results of ref. \cite{Zamo} for the boundary Ising model. Next, we formulate the ${\cal N}=1$ two-parametric boundary SShG and show that our results coincide with  those of ref. \cite{Nepo1} when  certain parameters are identified.   
In section 4 we present the B\"acklund solutions for one- and two-solitons and then derive the three-solitons solution.  In section 5  we investigate one-soliton and breather solutions in the half line and derive constraints on the boundary parameters. In appendix A, we give the explictly form of the three-soliton solution for the ${\cal N}=1$ SShG derived using B\"acklund transformations.


\section{Review of defect SShG model}
The ${\cal N}=1$ SShG model in the presence of defects \cite{Lean1} describing bosonic $\p_p$, and fermionic $\psi_p,\bpsi_p$, fields in the regions $x<0$ (corresponding to label $p=1$) and $x>0$ (corresponding to label $p=2$) respectively, can be described by the following Lagrangian density,\footnote{For simplicity we have used $m=1/2$.}
\begin{equation}
 {\mathcal L} = \theta(-x)  {\mathcal L}_1 + \theta(x) {\mathcal L}_2 +\delta(x) \ {\mathcal L}_D, \label{e2.1}
\end{equation}
where 
\begin{eqnarray}
 {\mathcal L}_p &=& \frac{1}{2}(\partial_x \phi_p)^2 - \frac{1}{2}(\partial_t \phi_p)^2 + \bar{\psi}_p(\partial_t -\partial_x)\bar{\psi}_p +   \psi_p(\partial_t +\partial_x)\psi_p  + V_p(\phi_p) \nonu \\ && + W_p(\phi_p,\psi_p, \bar{\psi}_p),\qquad\mbox{}\label{e2.2}\\[0.2cm]
   {\mathcal L}_D &=&\frac{1}{2}(\phi_2\partial_t\phi_1-\phi_1\partial_t\phi_2) -\psi_1\psi_2 -\bar{\psi}_1\bar{\psi}_2 + 2f_1\pa_t f_1 + B_0(\phi_1,\phi_2) \nonumber \\
   && + \,\,B_1(\phi_1,\phi_2,\psi_1,\psi_2,\bar{\psi}_1,\bar{\psi}_2,f_1),\quad \mbox{}\label{e2.3}
\end{eqnarray}
with the corresponding potentials given by
\begin{eqnarray}
 V_p&=& \cosh(2\phi_p),\qquad W_p = 4 \bar{\psi}_p \psi_p \cosh(\phi_p),\\
 B_0 &=& \s \cosh(\phi_1+\phi_2) +\frac{1}{\s}\cosh(\phi_1-\phi_2) , \label{e2.5}\\
 B_1 &=& -2\left[\sqrt{\sigma}\cosh\left(\frac{\phi_1+\phi_2}{2}\right)f_1(\bar{\psi}_1+\bar{\psi}_2) +\frac{1}{\sqrt{\sigma}}\cosh\left(\frac{\phi_1-\phi_2}{2}\right) f_1(\psi_1 -\psi_2)\right],\qquad \mbox{}
\end{eqnarray}
where $B_0$ and $B_1$ are the defect potentials. The equations of motion for the bulk are:
\begin{eqnarray}\label{mov1}
\partial_{x}^{2}\phi_{p}-\partial_{t}^{2}\phi_{p}&=&2\sinh(2\phi_{p})+4\,\bar{\psi}_{p}\psi_{p}\sinh\phi_{p},\nonumber\\
(\partial_{x}-\partial_{t})\bar{\psi}_{p}&=&2\,\psi_{p}\cosh\phi_{p},\nonumber\\
(\partial_{x}+\partial_{t})\psi_{p} &=&2\,\bar{\psi}_{p}\cosh\phi_{p}, \qquad p=1,2,
\end{eqnarray}
and the defect conditions at $x=0$ are given by
\begin{eqnarray}
\partial_{x}\phi_{1}-\partial_{t}\phi_{2} &=&-\frac{1}{\s}\sinh(\phi_{1}-\phi_{2})-\s\sinh(\phi_{1}+\phi_{2})+\sqrt{\s}\,\sinh\left(\frac{\phi_{1}+\phi_{2}}{2}\right)f_{1}(\bar{\psi}_{1}+\bar{\psi}_{2})\nonumber\\
&&+\frac{1}{\sqrt{\s}}\,\sinh\left(\frac{\phi_{1}-\phi_{2}}{2}\right)f_{1}(\psi_{1}-\psi_{2}).
\label{defeito1}
\end{eqnarray}
\begin{eqnarray} 
\partial_{x}\phi_{2}-\partial_{t}\phi_{1} & =& -\frac{1}{\s}\sinh(\phi_{1}-\phi_{2})+\s\sinh(\phi_{1}+\phi_{2})-\sqrt{\s}\,\sinh\left(\frac{\phi_{1}+\phi_{2}}{2}\right)f_{1}(\bar{\psi}_{1}+\bar{\psi}_{2})\nonumber\\
&& +\frac{1}{\sqrt{\s}}\,\sinh\left(\frac{\phi_{1}-\phi_{2}}{2}\right)f_{1}(\psi_{1}-\psi_{2}),
\label{defeito2}\\
\psi_{1}+\psi_{2} & = & \frac{2}{\sqrt{\s}}\,\cosh\left(\frac{\phi_{1}-\phi_{2}}{2}\right)f_{1},
\label{defeito3}\\
\bar{\psi}_{1}-\bar{\psi}_{2} & = & -2\sqrt{\s}\,\cosh\left(\frac{\phi_{1}+\phi_{2}}{2}\right)f_{1},
\label{defeito4}\\
\partial_{t}f_{1} &=&\frac{\sqrt{\s}}{2}\,\cosh\left(\frac{\phi_{1}+\phi_{2}}{2}\right)(\bar{\psi}_{1}+\bar{\psi}_{2}) +\frac{1}{2\sqrt{\s}}\,\cosh\left(\frac{\phi_{1}-\phi_{2}}{2}\right)(\psi_{1}-\psi_{2}).\quad \mbox{}
\label{defeito5} 
\end{eqnarray}
There is a fermionic degree of freedom $f_1$ at the boundary which anticommutes with the fields $\psi_p$ and $\bar{\psi}_p$. Notice that adding and subtracting the two first defect conditions (\ref{defeito1}) and (\ref{defeito2}) we get the following relations at $x=0$:
\begin{eqnarray}
 \pa_z(\p_1+\p_2) &=& -\frac{1}{\s}\sinh(\phi_{1}-\phi_{2}) +\frac{1}{\sqrt{\s}}\,\sinh\left(\frac{\phi_{1}-\phi_{2}}{2}\right)f_{1}(\psi_{1}-\psi_{2}),\label{back1}\\
 \pa_{\bar z}(\p_1-\p_2) &=& -\s\,\sinh(\phi_{1}+\phi_{2})+\sqrt{\s}\,\,\sinh\left(\frac{\phi_{1}+\phi_{2}}{2}\right)f_{1}(\bar{\psi}_{1}+\bar{\psi}_{2})\label{back2},
\end{eqnarray}
where we have used the light-cone coordinates $z=x-t$ and ${\bar z}=x+t$. In the general case where eqs. (\ref{back1}) and (\ref{back2}) held not only for $x=0$, but for every $x$, these are exactly the B\"acklund transformations for the supersymmetric sinh-Gordon model \cite{ChaiKul}.


\subsection{On-shell supersymmetry of the ${\cal N}=1$ defect SShG}

In the bulk, the supersymmetry transformations that leave the action of the ${\cal N}=1$ SShG  invariant are given by
\br
 \d\p &=& \vep\,\psi+\bvep\,\bpsi, \label{3.1}\\
 \d\psi &=& -\vep \,\pa_z \p + \bvep\sinh\p,\\
 \d\bpsi &=& \bvep \,\pa_{\bar z}\p - \vep\, \sinh\p ,  \label{3.3}
\er
where we are using the light-cone coordinates $z,{\bar z}$, and $\vep$, $\bvep$ are the fermionic parameters. The corresponding conserved supercharges in the bulk theory are given by, 
\br
 Q_{\vep} &=&- \int_{-\infty}^{\infty}dx\,\(2\psi\,\pa_z\phi + 2\bpsi\sinh\p\),\quad \quad
 \bar{Q}_{\bvep} =\int_{-\infty}^{\infty}dx\,\(2\bpsi\,\pa_{\bar{z}}\phi + 2\psi\sinh\p\).\quad \mbox{}\label{e3.5}
\er
However,  by considering the defect theory we found an additional contribution coming from the defect conditions at the origin. Then, the modified conserved supercharges take the form
\br
 {\cal Q} &=& Q_\vep + Q_{\mbox{\tiny D}}, \qquad \mbox{and} \qquad \bar{\cal Q} = \bar{Q}_{\bvep} + \bar{Q}_{\mbox{\tiny D}},
\er
where
 \br
 Q_{\mbox{\tiny D}} &=& \frac{4}{\sqrt{\s}}\left[\sinh\(\frac{\p_1-\p_2}{2}\)\,f_1 \right]\Bigg|_{x=0}, \quad \mbox{and} \quad  \bar{Q}_{\mbox{\tiny D}}= 4\sqrt{\s} \left[\sinh\(\frac{\p_1+\p_2}{2}\)\,f_1 \right]\Bigg|_{x=0}.\qquad \mbox{}\label{e3.7}
\er
Here, as a consequence of the supersymmetry transformations, we have used the corresponding supersymmetry transformation of the auxiliary fermionic field $f_1$ given by,
\br
 \d f_1 &=& \vep \left[\frac{1}{\sqrt{\s}}\sinh\(\frac{\p_1-\p_2}{2}\) \right] + \bvep \left[\sqrt{\s}\sinh\(\frac{\p_1+\p_2}{2}\) \right].
\er
Then, it worth noting that the fields not only can exchange momentum and energy with the defect as it was mentioned in \cite{Lean1}, but also can exchange supercharge.


\section{Half-line limiting procedure}

\subsection{Obtaining the boundary sinh-Gordon potential}

Firstly, we consider the bosonic part of the defect potential $B_0$ given in (\ref{e2.5}), 
\br
 B_0 &=& \s \cosh(\phi_1+\phi_2) +\frac{1}{\s}\cosh(\phi_1-\phi_2),
\er
and we perform the half-line limit by taking $\p_2$ to be a constant $k_{\mbox{\tiny $0$}}$ at $x=0^+$. It is convenient to redefine the B\"acklund parameter as $\s = e^{-\eta}$, to obtain
\br 
 \tilde{B}_0 &=& \(2\cosh\eta \cosh k_{\mbox{\tiny $0$}} \)\cosh \p_1 - 2\(\sinh\eta  \sinh  k_{\mbox{\tiny $0$}} \)\sinh\p_1.
\er
This potential has exactly the form of the boundary potential for the sinh-Gordon model given in (\ref{e1.1}) if the following relations hold,
\br
 \L \cosh \p_0 &=& 2 \cosh\eta\cosh  k_{\mbox{\tiny $0$}}, \label{e3.11}\\
 \L \sinh \p_0  &=&  2 \sinh\eta \sinh  k_{\mbox{\tiny $0$}}.\label{e3.12}
\er
From these relations we immediately note that the purely bosonic part of the defect Lagrangian in (\ref{e2.3}) becomes the boundary potential in (\ref{e1.1}) for the sinh-Gordon model defined upon the half-line, namely,
\br
 {\cal L}_D\bigg|_{\p_2= k_{\mbox{\tiny $0$}},\psi_2=\bpsi_2=0} &=& \L \cosh\(\p_1-\p_0\),
\er
with the corresponding boundary condition,
\br
 \pa_x\p_1\big|_{x=0} &=& -\L \sinh\(\p_1-\p_0\),
\er
where the parameters $\L$ and $\p_0$ are determined in terms of the B\"acklund parameter $\s$ and the constant $k_0$ by the inverse relations of the (\ref{e3.11}) and (\ref{e3.12}), as follows
\br
 \L^2 &=& \(\s^2+ \frac{1}{\s^2}\) + 2 \cosh(2 k_{\mbox{\tiny $0$}}),\\
 \tanh\p_0 &=& \(\frac{1-\s^2}{1+\s^2}\)\tanh k_{\mbox{\tiny $0$}}.
\er
We can also notice that if we exploit the symmetry of the sinh-Gordon equation under $\phi \to \pm \phi$ when $x\to -x$, we could take $\p_2(x) = \pm \p_1(-x)$ and in this case we obtain the following boundary potential,
\br
 B_0 &=&  \s^{\mp 1} + \s^{\pm 1}\,\cosh\(2\p_1\)\big|_{x=0} ,
\er
with the boundary condition,
\br
 \pa_x \p_1 \big|_{x=0} &=&  -\s^{\pm 1}\,\sinh\(2\p_1\)\big|_{x=0},
\er
which corresponds to the usual boundary problem when $\p_0 =0$ or equivalently $ k_{\mbox{\tiny $0$}}=0$. This fact is nothing more than the breaking of the discrete symmetry $\p\to-\p$ by the boundary potential. The properties of the solutions for the sinh-Gordon equation for a  free ($\pa_x\p_1\big|_{x=0} =0$) and  a fixed ($\p_1\big|_{x=0} =0$) boundary condition were already studied in \cite{Leonardis}.


\subsection{Deriving the boundary fermionic free field theory - Ising Model}

As it was discussed in \cite{Lean1}, integrable defect fermionic free field theories can be considered when all the bosonic fields vanish in the Lagrangian density (\ref{e2.1})-(\ref{e2.3}), namely,
\begin{eqnarray}
  {\mathcal L}_p &=& \bar{\psi}_p(\partial_t -\partial_x)\bar{\psi}_p +\psi_p(\partial_t +\partial_x)\psi_p  + 4\bar{\psi}_p{\psi}_p,\qquad p=1,2,\label{e4.9}\\[0.2cm]
   {\mathcal L}_D &=& -\psi_1\psi_2 -\bar{\psi}_1\bar{\psi}_2 + 2f_1\pa_t f_1 + 2\left[\sqrt{\sigma}(\bar{\psi}_1+\bar{\psi}_2) +\frac{1}{\sqrt{\sigma}}(\psi_1 -\psi_2)\right]f_1.\label{e4.10}
\end{eqnarray}
Then, the fields equations in the regions $x>0$ and respectively $x<0$ are simply given by,
\br
 \(\pa_x +\pa_t\)\psi_p &=& 2 \bar{\psi}_p, \qquad 
 \(\pa_x -\pa_t\)\bar{\psi}_p = 2 \psi_p,  \qquad p=1,2,\label{e4.11}
\er
and the defect conditions at $x=0$ are given by
\br
 \psi_1 +\psi_2 &=& \frac{2}{\sqrt{\s}}\,f_1, \qquad \bar{\psi}_1 - \bar{\psi}_2 \,=\, -2\sqrt{\s}\,f_1, \label{e4.13}\\
 \pa_t f_1 &=& \frac{\sqrt{\s}}{2}(\bar{\psi}_1+\bar{\psi}_2) +\frac{1}{2\sqrt{\sigma}}(\psi_1 -\psi_2).\label{e4.14}
\er
Now, we can notice that equations (\ref{e4.11}) are invariant under a duality transformation \mbox{$\psi_p \to -\bar{\psi}_p$} and $\bar{\psi}_p \to \psi_p$ when $x\to -x$. So, we use this symmetry to reduce the defect theory to a suitable boundary theory. Here, we will perform the half-line limit taking $\psi_2(0^+,t) \to -\bar{\psi}_1(0^-,t)$ and $\bar{\psi}_2(0^+,t) \to \psi_1(0^-,t)$. Then, the defect Lagrangian (\ref{e4.10}) becomes
\br
  {\mathcal L}_D &=& \psi_1(x)\bar{\psi}_1(0) - \bpsi_1(x)\psi_1(0) + 2f_1(t)\pa_t f_1(t) + 2\Big(\sqrt{\sigma}+\frac{1}{\sqrt{\sigma}}\Big)(\psi_1(x)+\bar{\psi}_1(x))f_1(t),\label{e4.15}\qquad\,\, \mbox{}
\er
with the corresponding boundary conditions,
\br
 \(\psi_1 - \bar{\psi}_1\)\bigg|_{x=0} &=& 2\(\sqrt{\s}+\frac{1}{\sqrt{\s}}\)f_1, \qquad \pa_t f_1\,=\, \frac{1}{2}\(\sqrt{\s}+\frac{1}{\sqrt{\s}}\)(\psi_1 +\bar{\psi}_1)\bigg|_{x=0},\qquad \mbox{}\label{e4.16}
\er
which can be combined to obtain a more compact form of the boundary condition for the fermionic fields $\psi_1$ and $\bar{\psi}_1$, as follows
\br
 \frac{d}{dt}\(\psi_1 - \bar{\psi}_1\)\bigg|_{x=0} &=& \(\sqrt{\s}+\frac{1}{\sqrt{\s}}\)^2(\psi_1 +\bar{\psi}_1)\bigg|_{x=0}.\label{e4.17}
\er
These results are exactly the ``boundary magnetic field" conditions  for the Ising model derived in  \cite{Zamo}, where the non-zero external field $h$ is related to the parameter $\s$, if we define
\br
 h &=& \(\sqrt{\s} +\frac{1}{\sqrt{\s}}\). \label{e4.21}
\er
From (\ref{e4.16}) we immediately can notice that $h=0$ or equivalently $\s= -1$ correspond to the ``free" boundary condition, which is written as
\br
\(\psi_1 - \bar{\psi}_1\)\bigg|_{x=0} &=& 0.\label{e4.22}
\er
On the other hand, in the limit $|h|\to \infty$ or $|\s| \to \infty$ one recovers the ``fixed" boundary condition, which becomes
\br
\(\psi_1 + \bar{\psi}_1\)\bigg|_{x=0} &=& 0.\label{e4.22}
\er


\subsection{Two-parametric boundary SShG model}

Now we are interested in performing the half-line limiting procedure on the ${\cal N}=1$ SShG model. To do that, we first perform the limit $\phi_2(0^+,t)\to  k_{\mbox{\tiny $0$}}$ for obtaining the boundary potential (\ref{e1.1}), and then to obtain the corresponding boundary fermionic free field theory we need to map $\psi_2(0^+,t) \to -\bpsi_1(0^-,t)$ and $\bpsi_2(0^+,t) \to \psi_1(0^-,t)$. So, by considering the part of the defect potential containing fermions, namely
\br
  B_1 &=& -2\left[\sqrt{\sigma}\cosh\left(\frac{\phi_1+\phi_2}{2}\right)f_1(\bar{\psi}_1+\bar{\psi}_2) +\frac{1}{\sqrt{\sigma}}\cosh\left(\frac{\phi_1-\phi_2}{2}\right) f_1(\psi_1 -\psi_2)\right],\quad \mbox{}
\er
and performing the limiting procedure, the Lagrangian (\ref{e2.3}) immediately becomes,
\br
 {\cal L}_B &=& \s \cosh\(\phi_1(x)+ k_{\mbox{\tiny $0$}}\) +\frac{1}{\s}\cosh\(\phi_1(x)- k_{\mbox{\tiny $0$}}\) +(\psi_1(x)\bar{\psi}_1(0) - \bpsi_1(x)\psi_1(0) ) + 2f_1\pa_tf_1\nonu \\
 && + 2\left[\sqrt{\sigma}\cosh\left(\frac{\phi_1(x)+ k_{\mbox{\tiny $0$}}}{2}\right) +\frac{1}{\sqrt{\sigma}}\cosh\left(\frac{\phi_1(x)- k_{\mbox{\tiny $0$}}}{2}\right) \right] (\bpsi_1(x)+\psi_1(x)) f_1,
\er
where the parameters $\s$ and $ k_{\mbox{\tiny $0$}}$ still satisfy the relations (\ref{e3.11}) and (\ref{e3.12}). This boundary Lagrangian density gives the following boundary conditions at $x=0$,
\br
 \pa_x\p_1 &=& -\s\sinh\(\p_1+ k_{\mbox{\tiny $0$}}\)-\frac{1}{\s}\sinh\(\p_1- k_{\mbox{\tiny $0$}}\)  - \sqrt{\sigma}\sinh\left(\frac{\phi_1+ k_{\mbox{\tiny $0$}}}{2}\right)(\bar{\psi}_1+\psi_1)f_1 \nonumber \\
                   &\mbox{}& -\,\frac{1}{\sqrt{\sigma}}\sinh\left(\frac{\phi_1- k_{\mbox{\tiny $0$}}}{2}\right) (\bpsi_1+\psi_1) f_1, \label{e4.48}\\
 \psi_1 -\bpsi_1 &=&  2\left[\sqrt{\sigma}\cosh\left(\frac{\phi_1+ k_{\mbox{\tiny $0$}}}{2}\right) +\frac{1}{\sqrt{\sigma}}\cosh\left(\frac{\phi_1- k_{\mbox{\tiny $0$}}}{2}\right) \right] f_1, \label{e4.49}\\
 \pa_t f_1 &=& \frac{1}{2}\left[\sqrt{\sigma}\cosh\left(\frac{\phi_1+ k_{\mbox{\tiny $0$}}}{2}\right) +\frac{1}{\sqrt{\sigma}}\cosh\left(\frac{\phi_1- k_{\mbox{\tiny $0$}}}{2}\right) \right] (\bpsi_1+\psi_1).\label{e4.50}
\er
which can be rewritten by eliminating $f_1$ as follows,
\br
 \pa_x\p_1 &=& -\s\sinh\(\p_1+ k_{\mbox{\tiny $0$}}\)-\frac{1}{\s}\sinh\(\p_1- k_{\mbox{\tiny $0$}}\)  -H(\p_1)(\bpsi_1 \psi_1),\qquad \,\,\,\mbox{}\label{e4.51}\\
 \pa_t(\psi_1-\bpsi_1)\big|_{x=0} &=& \frac{1}{2}H(\p_1)(\pa_t\p_1)(\psi_1-\bpsi_1) + 
 h^2(\p_1) (\psi_1+\bpsi_1),\label{e4.52}
\er
where
\br
 h(\p_1)&=& \left[\sqrt{\sigma}\cosh\left(\frac{\phi_1+ k_{\mbox{\tiny $0$}}}{2}\right) +\frac{1}{\sqrt{\sigma}}\cosh\left(\frac{\phi_1- k_{\mbox{\tiny $0$}}}{2}\right) \right], \label{equ3.30}\\[0.2cm]
 H(\p_1) &=& \left[\frac{\sigma\sinh\left(\frac{\phi_1+ k_{\mbox{\tiny $0$}}}{2}\right) +\sinh\left(\frac{\phi_1- k_{\mbox{\tiny $0$}}}{2}\right) }{\s\cosh\left(\frac{\phi_1+ k_{\mbox{\tiny $0$}}}{2}\right) +\cosh\left(\frac{\phi_1- k_{\mbox{\tiny $0$}}}{2}\right)}\right].\label{bc4.31}
\er
By applying the supersymmetry transformations (\ref{3.1})--(\ref{3.3}) with $\vep=\bvep$, we obtain
\br
  \d(\psi_1-\bpsi_1) &\equiv & \vep\[ -(\pa_x\p_1) + 2 \sinh \p_1\] , \nonu\\
  &=& \!\!\left[\sqrt{\sigma}\sinh\Big(\frac{\phi_1+k_{\mbox{\tiny $0$}}}{2}\Big) +\frac{1}{\sqrt{\sigma}}\sinh\Big(\frac{\phi_1-k_{\mbox{\tiny $0$}}}{2}\Big) \right] \!\vep(\psi_1+\bpsi_1)f_1 + 2\,\vep\, h(\p_1) F_1.\,\,\,\qquad \mbox{}
\er
Then, using eq.(\ref{e4.48}) we get
\br
 F_1 &=& \sqrt{\sigma}\sinh\left(\frac{\phi_1+k_{\mbox{\tiny $0$}}}{2}\right) +\frac{1}{\sqrt{\sigma}}\sinh\left(\frac{\phi_1-k_{\mbox{\tiny $0$}}}{2}\right).
\er
It is worth noting that $F_1 = 2 h'(\p_1)$ with $h'(\p_1) = \frac{dh(\p_1)}{d\p_1}$. Now, varying eq.(\ref{e4.50}), we immediately find for the lhs,
\br
 \d\(\pa_t f_1\) = \vep\,\pa_t\left[\sqrt{\sigma}\sinh\left(\frac{\phi_1+k_{\mbox{\tiny $0$}}}{2}\right) +\frac{1}{\sqrt{\sigma}}\sinh\left(\frac{\phi_1-k_{\mbox{\tiny $0$}}}{2}\right)\right]\, =\, \frac{\vep}{2}\, h(\p_1) (\pa_t \p_1),
\er
and from the rhs we get,
\br
\d\(\frac{1}{2}h(\p_1)(\psi_1+\bpsi_1)\) &=& \frac{\vep}{2} \,h(\p_1) (\pa_t \p_1),
\er
This shows that the eq.(\ref{e4.50}) is invariant under the supersymmetry transformations. It remains to show that eq.(\ref{e4.48}) is also invariant. To do that, we consider the lhs of (\ref{e4.48}), 
\br
 \d\(\pa_x \p_1\) &=& \vep\,\pa_x(\psi_1+\bpsi_1) \,=\,\vep\left[\pa_t(\bpsi_1- \psi_1) + 2(\psi_1+\bpsi_1)\cosh\p_1 \right] \nonumber \\[0.1cm]
 &=&  -\vep \left[\sqrt{\sigma}\sinh\left(\frac{\phi_1+k_{\mbox{\tiny $0$}}}{2}\right) +\frac{1}{\sqrt{\sigma}}\sinh\left(\frac{\phi_1-k_{\mbox{\tiny $0$}}}{2}\right) \right](\pa_t\p_1)f_1 + 2\vep\(\psi_1+\bpsi_1\)\cosh\p_1\nonumber \\
 && -\vep\left[\sqrt{\sigma}\cosh\left(\frac{\phi_1+k_{\mbox{\tiny $0$}}}{2}\right) +\frac{1}{\sqrt{\sigma}}\cosh\left(\frac{\phi_1-k_{\mbox{\tiny $0$}}}{2}\right) \right]^2\(\psi_1+\bpsi_1\).\label{e4.55}
\er
From the rhs of (\ref{e4.48}) we obtain,
\br
 \d(\pa_x\p_1) &=& -\vep\left[\s\cosh\(\p_1+k_{\mbox{\tiny $0$}}\) + \frac{1}{\s} \cosh\(\p_1-k_{\mbox{\tiny $0$}}\)\](\psi_1+\bpsi_1)\nonumber \\ && -\vep\left[\sqrt{\sigma}\sinh\left(\frac{\phi_1+k_{\mbox{\tiny $0$}}}{2}\right) +\frac{1}{\sqrt{\sigma}}\sinh\left(\frac{\phi_1-k_{\mbox{\tiny $0$}}}{2}\right) \right](\pa_t\p_1)f_1 \nonumber\\
 && + \vep\,\left[\sqrt{\sigma}\sinh\left(\frac{\phi_1+k_{\mbox{\tiny $0$}}}{2}\right) +\frac{1}{\sqrt{\sigma}}\sinh\left(\frac{\phi_1-k_{\mbox{\tiny $0$}}}{2}\right)\right]^2(\psi_1+\bpsi_1).\label{e4.56}
\er
It is not so difficult to show that variations (\ref{e4.55}) and (\ref{e4.56}) are actually the same. Now, if we compute the boundary contribution to the supercharge, we find that the modified boundary supercharge can be written as ${\hat Q} = Q_+ + Q_B $, where 
\br
 Q_{+} &=& \int_{-\infty}^{0}dx\,\left[ (\bpsi_1-\psi_1)(\pa_x\p_1-2\sinh\p_1)  + (\bpsi_1+\psi_1) \pa_t\p_1 \right] ,
\er
and
\br
 Q_B &=& 4\left[\sqrt{\sigma}\sinh\left(\frac{\phi_1+k_{\mbox{\tiny $0$}}}{2}\right) +\frac{1}{\sqrt{\sigma}}\sinh\left(\frac{\phi_1-k_{\mbox{\tiny $0$}}}{2}\right) \right]f_1.
\er
If we compare this result with eq.(\ref{e3.7}), we have that $Q_B = Q_D\big|_{\vep=\bvep,\p_2=k_0}$ as it should be expected.
We also remark that this modified boundary supercharge has a different functional form from that was previously derived in \cite{Nepo1}, namely,
\br
 \S_{1/2} &=& \frac{i}{2} \left[\frac{\a \sin\(\frac{\p-\p_0}{2}\)    - 4\sin (\frac{\p}{2}) } {f(\p) }\right] a,
\er
where $a$ is related with the fermionic degree of freedom $f_1$ used in this paper, and
\br
f(\p) =\frac{\sqrt{C}}{2} \sin\(\frac{\p-D}{4}\) ,\label{equ3.41}
\er
\vspace{-0.6cm}
\br
 C = \sqrt{\a^2-8\a\cos\Big({\p_0\o 2}\Big) +16}, \qquad \tan\Big({D\o 2}\Big) = \frac{\a \sin\(\p_0 \o 2\)}{\a\cos\(\p_0 \o 2\) -4}.
\er
However, after performing the respective analytical continuation, namely,
\br
\p &\longrightarrow& 2i\p_1, \qquad \psi \longrightarrow 2i\psi_1,  \qquad \bpsi \longrightarrow -2i\bpsi, \qquad a \longrightarrow 2\sqrt{2}i f_1, \qquad k_{\mbox{\tiny $0$}}\longrightarrow -ik_{\mbox{\tiny $0$}}, \qquad \mbox{}
\er
it is possible to show the equivalence of both results by the following identification,
\br
  C &=& 4\(\cos k_{\mbox{\tiny $0$}} +\cosh \eta\), \quad \,\, D\,=\, 4\,\arctan\[-\coth\(\eta \o 2\)\cot\({k_{\mbox{\tiny $0$}} \o 2}\) \], \\
\a &=& -2\L, \qquad  f(\p)\Big|_{\p=2i\p_1} = \frac{1}{\sqrt{2}}\,h(\p_1), \qquad \,\quad \S_{1/2} = 2i Q_{B}.
\er

\section{B\"acklund Solutions}
\label{bac}

We will present in this section soliton solutions for the ${\cal N}=1$ SShG based upon the results obtained in \cite{LeanN1}. Consider the bosonic superfield written in components as
\begin{eqnarray}
\P &=& \p + \th_1 \bpsi + i\th_2\psi -i\th_1\th_2 \sinh\p,
\end{eqnarray}
where $\th_1, \th_2$ are Grassmannian variables. Then, starting from a vacuum configuration $\P_0=0$, the one-soliton solution has the form
\br
 \p^{(k)}&=& 2 \arctanh\[a_k e^{\G_k}\], \qquad \qquad \quad \,\, \,\, \G_k = (\s_k+\s_k^{-1})x +(\s_k-\s_k^{-1})t ,\\
 \bpsi^{(k)} &=& 2\eps_k \,\(\frac{b_k}{a_k}\)\frac{a_k\, e^{\G_k}}{1-(a_k\,e^{\G_k})^{2}}, \qquad \quad  \psi^{(k)} \,=\, \frac{\bpsi^{(k)}}{\s_k},
\er
where $\s_k$ is the B\"acklund parameter, $a_k, b_k$ are arbitrary parameters and $\eps_k$ being a Grassmannian parameter.
The two-soliton solutions $\P^{(j,k)}$ is explicitly given in components by,
\br
 \p^{(j,k)} &=& \vp^{(j,k)} - f^{(j,k)} \bpsi^{(j)}\bpsi^{(k)}\label{2.15}\\
 \bpsi^{(j,k)} &=& \xi^{(j,k)}\bpsi^{(j)} + \xi^{(k,j)}\bpsi^{(k)} \\
 \psi^{(j,k)} &=& \eta^{(j,k)}\psi^{(j)} + \eta^{(k,j)}\psi^{(k)},
\er
with{\small
\br
\vp^{(j,k)} &=& 2\,\mathrm{Arctanh}\left[\delta_{jk}\tanh\Bigg(\frac{\phi^{(j)}-\phi^{(k)}}{2}\Bigg)\right], \quad
 f^{(j,k)} = \frac{\Delta_{jk}}{4\sqrt{\sigma_j\sigma_k}}\left[\mathrm{sech}\Bigg(\frac{\phi^{(j)}}{2}\Bigg)\mathrm{sech}\Bigg(\frac{\phi	^{(k)}}{2}\Bigg)\right],\qquad\mbox{}\label{feq5.18}\\
 \Omega_{jk}  &=&  \frac{\delta_{jk}\,\mathrm{sech}^2\left(\frac{\phi^{(j)}-\phi^{(k)}}{2}\right)}{1-\delta_{jk}^2\tanh^2\left(\frac{\phi^{(j)}-\phi^{(k)}}{2}\right)},\qquad \qquad\quad\,\,\,\,\,\,\,
\Delta_{jk}\,=\,\frac{A_{jk}\sinh\left(\frac{\phi^{(j)}-\phi^{(k)}}{2}\right)}{B_{jk}-\sinh^2\left(\frac{\phi^{(j)}-\phi^{(k)}}{2}\right)},\label{2.21}\\ 
 \xi^{(j,k)}  &=&  \O_{jk} + \frac{\Delta_{jk}}{2}\sqrt{\frac{\sigma_k}{\sigma_j}}\,\frac{\sinh\Big(\frac{\phi^{(k)}}{2}\Big)}{\cosh\Big(\frac{\phi^{(j)}}{2}\Big)},\qquad \qquad\,\,\, \eta^{(j,k)}=\O_{jk} - \frac{\Delta_{jk}}{2}\sqrt{\frac{\sigma_j}{\sigma_k}}\,\frac{\sinh\Big(\frac{\phi^{(k)}}{2}\Big)}{\cosh\Big(\frac{\phi^{(j)}}{2}\Big)},\label{feq5.19}\\
 \delta_{jk}&=&\frac{\sigma_j+\sigma_k}{\sigma_j-\sigma_k}, \qquad \qquad 
A_{jk}=\frac{\sigma_j+\sigma_k}{\sqrt{\sigma_j\sigma_k}}, \qquad\,\,\,
B_{jk}=\frac{(\sigma_j-\sigma_k)^2}{4\,\sigma_j\sigma_k}.
\er}
It therefore follows that starting from the one-soliton solution $\P^{(2)}$ and the two-solition solutions $\P^{(1,2)}$ and $\P^{(2,3)}$, we can construct the three-soliton solution $\P_3\equiv \P^{(1,2,3)}$ using appropriately the B\"acklund procedure as indicated in (\ref{Fig A.1}). Then the three-soliton solution $\P_3\equiv \P^{(1,2,3)}$ can be written in components as follows, 
\br
\p_3 &=& \p^{(2)} + 2\arctanh\left(\d_{13}\tanh \t_3\) - \rho_{12}\bpsi^{(1)}\bpsi^{(2)} +\rho_{13}\bpsi^{(1)}\bpsi^{(3)} -\rho_{23}\bpsi^{(2)}\bpsi^{(3)},\qquad\,\, \mbox{}\\[0.1cm]
\bpsi_3 &=& \chi_1 \bpsi^{(1)} +  \chi_2 \bpsi^{(2)} + \chi_3\bpsi^{(3)} + \chi_4 (\bpsi^{(1)}\bpsi^{(2)}\bpsi^{(3)}),\\[0.1cm]
\psi_3 &=& \mu_1\psi^{(1)} +  \mu_2 \psi^{(2)} + \mu_3\psi^{(3)} + \mu_4 (\psi^{(1)}\psi^{(2)}\psi^{(3)}),
\er
where for convenience we have defined $\t_3 = (\vp^{(1,2)}- \vp^{(2,3)})/2$, and the expressions for $\rho_{ij}$, $\chi_{i}$ and $\mu_{i}$ are explicitly given in the appendix \ref{appA}.

\section{Classical analysis of the solutions in the half line}

\subsection{One-soliton solution}

In contrast with the sine-Gordon model which has a degenerated vacuum, the sinh-Gordon counterpart has only one  vacuum which satisfy
\br
 \lim_{x\to -\infty} \p(x,t) &=& 0.
\er
Let us consider a ground state realized by a static bulk one-soliton ($\s_1=1$) for $x<0$ with the following form,
\br
 \p(x)&=& 2\arctanh[a\, e^{2x}]. \label{eq6.2}
\er
Then, the boundary condition (\ref{e4.51}) determines the parameter $a$ in terms of the boundary parameters $\s$ and $k_{\mbox{\tiny $0$}}$ as follows,
\br
 a &=& \left[\tanh\(\frac{\eta}{2}\)\tanh\(\frac{k_{\mbox{\tiny $0$}}}{2}\)\]^{\mp 1},
\er
where we have defined $\s =e^{-\eta}$ and the power $\mp 1$ depends on the signal of the term $|\sinh k_{\mbox{\tiny $0$}} \sinh \eta|$. We can notice that this result has already been found for the case of sine-Gordon \cite{Bajnok1}, where the authors studied the bound state spectrum of the model with the integrable boundary condition.

Now consider the static bulk one-soliton solution for the fermionic fields,
\br
 \bpsi(x) &=&  \[\frac{2\,b \,e^{2x}}{1-a^2 e^{4x}}\]\epsilon , \qquad \psi(x) = \bpsi(x),
\er
where $\epsilon$ is the Grassmannian parameter. From the boundary condition (\ref{e4.52}) we immediately get,
\br
 b= 0, \qquad \mbox{or} \qquad h(\p) \,=\, \sqrt{\s}\cosh\(\frac{\p_1+k_{\mbox{\tiny $0$}}}{2}\) + \frac{1}{\sqrt{\s}}\cosh\(\frac{\p_1-k_{\mbox{\tiny $0$}}}{2}\) = 0,
\er
which implies that,
\br
 b =0, \qquad \mbox{or} \qquad a\,=\,\left[\tanh\(\frac{\eta}{2}\)\tanh\(\frac{k_{\mbox{\tiny $0$}}}{2}\)\]^{-1}.
\er
The first contraint reduces the problem to the purely bosonic case. From the second constraint, we can note that the boundary conditions coming from the fermionic part are removing the ambiguity of the power and determining completely the relation between the parameter $a$ and the boundary parameter $\s$ and $k_{\mbox{\tiny $0$}}$. In other words, this ambiguity is solved with the use of the discrete (\emph{duality}) symmetry for the fermionic fields. However, as in the bulk sinh-Gordon model the energy for the one-soliton solution does not change because of any fermionic contribution.


\subsection{Breather Solutions}
The breather solution is obtained by taking a two-soliton solution with the appropriate choice of the parameters. Let us consider the bosonic part of the two-soliton solution
\br
 \vp^{(1,2)} &=& 2 \arctanh\[\(\frac{\s_1+\s_2}{\s_1-\s_2}\) \frac{a_1 e^{\G_1} -a_2 e^{\G_2}}{1-a_1a_2 e^{\G_1+\G_2}}\],
\er
where $\G_i = \(\s_i +\s_i^{-1}\)x + \(\s_i -\s_i^{-1}\)t$. An appropriate choice of the parameters is the following,
\br 
 \s_1 &=& e^{i\zeta}, \quad \s_2^* = \s_1, \quad a_1 = -ie^{\a_1}, \quad a_2 = a_1^*. \label{beq 6.8}
\er
From these, we get $\G_1^*=\G_2$, with
\br
 \G_1 &=& \kappa x +i\om t,  \qquad \kappa =2 \cos \zeta, \qquad \om =2  \sin\zeta, \quad -\frac{\pi}{2}\leq \zeta \leq \frac{\pi}{2}.
\er
Then, we found the following form for the breather solution,
\br
 \vp^{(1,2)} &=& 2 \arctanh\[\frac{\sqrt{4-\om^2}}{\om}\,\frac{\cos \om t}{\sinh(\kappa x +\a_1)} \].
\er
Notice that it can also be rewritten as,
\br
  \vp^{(1,2)} &=& \ln\[\frac{1   - e^{4x\cos\zeta +2\a_1} -\frac{2 e^{\a_1}}{\tan\zeta} \cos(2t\sin\zeta)e^{2x\cos\zeta} }{1 - e^{4x\cos\zeta +2\a_1} +\frac{2 e^{\a_1}}{\tan\zeta} \cos(2t\sin\zeta)e^{2x\cos\zeta} }\].\label{e6.2.5}
\er
The form of the breather solution (\ref{e6.2.5}) has been already used for investigating the energy spectrum of the boundary states \cite{CorrDelius}. Now, we will investigate the constraints that the boundary condition impose over the parameter. So, for the bosonic part we have,
\br
 \pa_x  \vp^{(1,2)}\big|_{x=0}  &=& \left[\L_+ \sinh \vp^{(1,2)} +\L_- \cosh \vp^{(1,2)}\]\bigg|_{x=0},\label{e6.2.6}
\er
with
\br
 \L_+ &=& -\(\s+\frac{1}{\s}\)\cosh k_{\mbox{\tiny $0$}}, \qquad \L_- \,=\,  \(\frac{1}{\s}-\s\)\sinh k_{\mbox{\tiny $0$}}.
\er
By evaluating the breather solution (\ref{e6.2.5}) in the boundary condition (\ref{e6.2.6}), we found that
\br
 \L_- &=& 0,\qquad  \L_+ =-\frac{2\cos\zeta}{\tanh\a_1}.
\er
From these constraints we immediately find that
\br
 \cos \z = \left\{ \begin{array}{c l} 
 \pm (\cosh k_{\mbox{\tiny $0$}})(\tanh \a_1), & \mbox{if \quad}\s =\pm 1, \\[0.2cm]
 \frac{(-1)^n}{2}\(\s+{1\o\s}\) \tanh\a_1, &\mbox{if \quad} k_{\mbox{\tiny $0$}} =in\pi, \quad n=0,1,2,\dots 
 \end{array} \right.
\er
Now, we consider the fermionic contribution to the bosonic solution (\ref{2.15}), namely $f^{(1,2)}$ given in (\ref{feq5.18}), which takes the following form
\vskip  -0.3cm
\br
f^{(1,2)} &=& -{{ i \, e^{2x\cos \zeta +\a_1}\cos\zeta \cos\(2t\sin\zeta\) \big[\cos\(4t\sin\zeta\) +\cosh\(4x\cos\zeta +2\a_1\)\big]      }  \o {\left[1 -\cosh \left(2 \alpha _1+4 x \cos \zeta \right)\] +\cos ^2\zeta  \left[\cos (4 t\sin\zeta ) +\cosh \left(2 \alpha _1+4 x \cos \zeta \right)\]} }.\qquad \mbox{}
\er
The boundary condition for this term of the solution (\ref{2.15}) can be written as follows,
\vskip  -0.3cm
\br
 \pa_x\Big[f^{(1,2)}\bpsi^{(1)}\bpsi^{(2)}\Big]_{x=0} &=& -\Big[\big(\s \cosh(\vp^ {(1,2)}+k_{\mbox{\tiny $0$}}) + \s^{-1} \cosh(\vp^ {(1,2)}-k_{\mbox{\tiny $0$}}) \big) f^{(1,2)}\bpsi^{(1)}\bpsi^{(2)}\Big]_{x=0} \nonumber \\
 && -\Big[ H(\vp^{(1,2)}) \big(e^{-i\zeta}\eta^{(1,2)}\xi^{(2,1)}- e^{i\zeta}\eta^{(2,1)}\xi^{(1,2)}\big)\bpsi^{(1)}\bpsi^{(2)}\Big]_{x=0},\label{bre6.16}
\er
where the form for the functions $H(\vp^{(1,2)})$, $\eta^{(j,k)}$ and $\zeta^{(j,k)}$ are given in (\ref{bc4.31}) and (\ref{feq5.19}) respectively. In addition, the $\eps_j$-projection of the fermionic boundary condition (\ref{e4.52}) takes the form
\br
\pa_t\Big[\(\s_j^{-1} \,\eta^{(j,k)} - \xi^{(j,k)}\)\bpsi^{(j)} \Big]_{x=0} &=& \Big[ \frac{1}{2}H(\vp^{(1,2)})\,\pa_t\vp^{(1,2)}\(\s_j^{-1} \,\eta^{(j,k)} - \xi^{(j,k)}\)\bpsi^{(j)} \nonumber \\
&& +h^2(\vp^{(1,2)}) \(\s_j^{-1} \,\eta^{(j,k)} + \xi^{(j,k)}\)\bpsi^{(j)}\Big]_{x=0},\label{bre6.17}
\er
for $j=1,2$. From the conditions (\ref{bre6.16}) and (\ref{bre6.17}) we get additional constraints, namely
\begin{itemize}
\item[$(i)$] If $\s =+1$, then $k_{\mbox{\tiny $0$}} = (2n+1)i\pi$ with $n=0,1,...$,  which implies that \mbox{$\cos\z= -\tanh\a_1$}. However,  $\z={n\pi \o 2}$ does not satisfy the conditions (\ref{bre6.16}) and (\ref{bre6.17}), and therefore  $\cos\z$ cannot take the values $\{0,\pm 1\}$.
\item[$(ii)$] If $\s=-1$, then $k_{\mbox{\tiny $0$}}=2ni\pi$ with $n=0,1,...$, which implies again that \mbox{$\cos\z= -\tanh\a_1$} and respectively $\z \neq  {n\pi \o 2}$. 
\end{itemize}
\vskip 0.2cm
\noindent Considering these results, the boundary breather at $x=0$ can be written as follows,
\br
  \vp^{(1,2)}(0,t)  &=& -2 \arctanh\[\cos\(2t\, \text{sech}\a_1\) \] ,\\[0.2cm]
  f^{(1,2)}(0,t) &= &-\frac{i}{2}e^{\a_1}\tanh\a_1\cot(2 t\, \text{sech}\alpha _1)\csc(2 t\, \text{sech}\alpha _1)   \left(\cosh 2 \alpha _1 +\cos(4 t \,\text{sech}\alpha _1)\right),\qquad\,\,\,\, \mbox{}\\[0.2cm]
  \psi^{(1,2)}(0,t) &=& -\left[\frac{\eps_1 b_1(e^{\alpha _1}+i)}{e^{\a_1} (e^{\alpha _1}-i)}\right] \left(1+(e^{\alpha _1}-i) \cot (2 t\, \text{sech}\alpha _1)\right) \csc (2 t\, \text{sech}\alpha _1)\\[0.2cm]
  && -\left[\frac{\eps_2 b_2(e^{\alpha _1}-i)}{e^{\a_1}(e^{\alpha _1}+i)}\right] \left(1+\left(e^{\alpha _1}+i\right) \cot (2 t\, \text{sech}\alpha _1)\right) \csc (2 t\, \text{sech}\alpha _1),\\[0.2cm]
\bpsi^{(1,2)}(0,t)&=&   -\left[ \epsilon_1 b_1 e^{-\a_1} \left((e^{\alpha _1}+i) \cot (2 t \text{sech}\alpha _1) -1\right) \csc (2 t\, \text{sech}\alpha _1)\] \nonumber \\[0.2cm]
&&  + \left[\epsilon_2 b_2 e^{-\a_1} \left((e^{\alpha _1}-i) \cot (2 t \text{sech}\alpha _1) -1\right) \csc (2 t\, \text{sech}\alpha _1)\].
\er

\subsection{Method of Images: Bosonic part}

In \cite{SSW} the classical and semi-classical soliton reflection on the boundary at $x=0$ for the boundary sine-Gordon model were investigated by using the method of images \cite{Cardy}. To do that, it is necessary to consider the three-soliton solution describing an incoming soliton, an outgoing soliton and a stationary soliton at the origin, choosing the rapidity parameters as follows,
\br
 \s_1 = e^{\th},\quad \s_2=e^{-\th}, \quad \s_3=1 ,
\er
to construct the appropriate three-soliton solution as follows,
\br
 \vp_3(x,t) = 2\, \arctanh\[\frac{N(x,t)}{D(x,t)}\], \label{e6.2}
\er
with
\br
 N(x,t) &=& \left[\frac{\cosh\th}{1-\cosh\th}\]e^{2x\cosh\th }\,F(t) - \left[\frac{1}{\tanh^2\th}\] e^{2x-\b}-e^{2x\(2\cosh\th+1\)-(\a+\b)},\\
 D(x,t) &=& 1 + \left[\frac{1}{\tanh^2\th}\right]e^{4x\cosh\th -\a} - \left[\frac{\cosh\th}{1-\cosh\th}\]e^{2x(\cosh\th+1) -\b}\,F(t) .
\er
where we have defined
\br
 F(t) &=& e^{2t\sinh\th -\a_1} +e^{-2t\sinh\th-\a_2}, \qquad a_i=e^{-\a_i}, \quad \a=\a_1+\a_2, \quad \mbox{and} \quad \b=\a_3. \qquad \mbox{}
\er
This solution describes an incoming right-moving soliton at $t\to -\infty$ and an outgoing soliton at $t\to \infty$. To see that explicitly, we take $x\to-\infty$ and $t\to-\infty$ with the phase $2(x\cosh\th +t\sinh\th)  $ fixed, leading to
\br
 \vp_3 \longrightarrow 2\arctanh\[\left(\frac{\cosh\th}{1-\cosh\th}\)\,e^{2\(x\cosh\th +t\sinh\th\) - \a_1}  \],
\er
and when we take $x\to-\infty$ and $t\to\infty$ we obtain,
\br
 \vp_3 \longrightarrow 2\arctanh\[\left(\frac{\cosh\th}{1-\cosh\th}\)\,e^{2\(x\cosh\th -t\sinh\th\) - \a_2}  \].
\er
Then, we have a reflected soliton with an opposite velocity and a phase delay $\a=\a_1+\a_2$ when $t\to\infty$ . Now, this phase delay can be computed using the boundary condition. For the three-soliton solution we can decompose the boundary condition (\ref{e4.51}) in four boundary equations. The first one corresponds to the bosonic part, which can be written as
\br
 \pa_x \vp_3 &=& -\s \sinh(\vp_3 + k _{\mbox{\tiny $0$}}) -\frac{1}{\s}\sinh(\vp_3 -k _{\mbox{\tiny $0$}}).\label{e6.8}
\er
Substituting (\ref{e6.2}) into (\ref{e6.8}), we get
\br
\left[ (\pa_x N)D- N(\pa_x D)\right]\big|_{x=0} &=& \left[\L_+ (N D) +\frac{\L_-}{2}\(N^2+D^2\)\right]\bigg|_{x=0},\label{e6.9}
\er
with
\br
 \L_+ &=& -2\cosh\eta\cosh k _{\mbox{\tiny $0$}}, \qquad \L_- \,=\,  2\sinh\eta\sinh k _{\mbox{\tiny $0$}}.
\er
where $\s=e^{-\eta}$. Then, the ansatz proposed in \cite{SSW} is used to solve the equation (\ref{e6.9}), namely
\br
 \pa_x N \big|_{x=0} &=& [c_1 N + c_2 D]\big|_{x=0}, \nonumber \\
 \pa_x D \big|_{x=0} &=& [c_3 N + c_4 D]\big|_{x=0}.
\er
From this we find that $c_2 = -c_3=\frac{\L_-}{2}$, and $c_1-c_4=\L_+$. By computing each constants $c_i$, we obtain
\br
\cosh\eta\cosh k _{\mbox{\tiny $0$}}  &=& -\cosh\th\left[\frac{1- \coth^2\(\th\o 2\) e^{-\a}+ \coth^2\(\th\o 2\) e^{-2\b}\left(1-e^{-\a}\tanh^2\(\th\o 2\)\right)   }{1+ \coth^2\(\th\o 2\) e^{-\a}- \coth^2\(\th\o 2\) e^{-2\b}\left(1+e^{-\a}\tanh^2\(\th\o 2\)\right)   }\right],\qquad \,\,\, \mbox{}\\[0.2cm]
\sinh\eta\sinh k _{\mbox{\tiny $0$}} &=& \frac{2\,e^{-\b}\(1+\cosh\th\)(1- e^{-\a})}{1+ \coth^2\(\th\o 2\) e^{-\a}- \coth^2\(\th\o 2\) e^{-2\b}\left(1+e^{-\a}\tanh^2\(\th\o 2\)\right)  }.\label{e6.12}
\er
\normalsize
From the above relations we can find that the ``position" parameter of the stationary soliton is given by 
\br
 \cosh \b &=& \frac{1-\cosh \eta \cosh k _{\mbox{\tiny $0$}}}{\sinh \eta \sinh k _{\mbox{\tiny $0$}}}.
\er
Then,
\br
 \sinh \b &=& \pm \left[\frac{\cosh k _{\mbox{\tiny $0$}} -\cosh \eta}{\sinh\eta \sinh k _{\mbox{\tiny $0$}}}\right], \qquad {\rm and}\quad e^{-\b} \,=\,-\left[ \frac{\tanh\(\frac{k _{\mbox{\tiny $0$}}}{2}\)}{\tanh\(\frac{\eta}{2}\)}\right]^{\pm 1} \,\equiv\, a_3^{\pm 1}.
\er
Using the above results we find that the phase delay $\a$ can be written as follows,
\br
 \a &=& \ln\left[\frac{\(\tanh^2\(\frac{\th}{2}\)-\tanh^2\(\frac{\eta}{2}\)\) \(1-\tanh^2\(\frac{\th}{2}\)\tanh^2\(\frac{k_0}{2}\)\)}{\(\tanh^2\(\frac{\th}{2}\)-\tanh^2\(\frac{k_0}{2}\)\) \(1-\tanh^2\(\frac{\th}{2}\)\tanh^2\(\frac{\eta}{2}\)\)} \right]^{\pm 1}\nonumber \\
&=& \ln\left[\frac{\tanh\(\frac{\th+\eta}{2}\)\tanh\(\frac{\th-\eta}{2}\)}{\tanh\(\frac{\th +k_0}{2}\)\tanh\(\frac{\th -k_0}{2}\)} \right]^{\pm 1}\nonumber \\
&=&\ln\[\frac{ (\cosh k _{\mbox{\tiny $0$}}+\cosh\th)(\cosh\eta -\cosh\th)}{(\cosh k _{\mbox{\tiny $0$}}-\cosh\th) (\cosh\eta+\cosh\th)}\]^{\pm 1}. \label{eq6.31}
\er
As it was noticed in \cite{SSW}, the $\pm 1$ power comes by solving a quadratic equation in (\ref{e6.12}), or equivalently in our case, from solving the square root to find $\sinh\b$. In addtion, the  argument of the logarithm in (\ref{eq6.31}) is positive only in the cases when $\th > k _{\mbox{\tiny $0$}}$ and $\th> \eta$, or $\th < k _{\mbox{\tiny $0$}}$ and $\th<\eta$. Then, there is a forbidden domain for the rapidity $\th$ consisting of all the values in the interval between $\eta$ and $k _{\mbox{\tiny $0$}}$.

Finally, it is worth discussing boundary breathers. To do that, it is convenient to choose the parameters in following way, 
\br
\s_1 &=& e^{i\zeta}, \quad \s_2^* = \s_1, \quad \s_3=1,\quad  a_1 = -ie^{\a_1}, \quad a_2 = a_1^*, \quad a_3 = e^{-\b}, \label{beq 6.32}
\er
which implies again,
\br
 \G_1= \G_2^* =\kappa x +i\om t, \,\,\,\quad \G_3 = 2x,  \qquad \,\,\, \kappa =2 \cos \zeta, \quad \om =2  \sin\zeta, \quad -\frac{\pi}{2}\leq \zeta \leq \frac{\pi}{2}.\qquad \mbox{}
\er
By substituting in the bosonic three-soliton solution we get the following form for the breather solution,
\br
\vp_{\mbox{\tiny $B$}} \!=\! 2\arctanh\Bigg[\frac{2 \k e^{\alpha_1 +\k x} \sin \om t +e^{2 x} \coth \left(\frac{\eta }{2}\right) \tanh \big(\frac{k _{\mbox{\tiny $0$}}}{2}\big) ((\k-2) e^{2 (\alpha_1 +\k x)}+(\k+2))}{2 \k \coth \left(\frac{\eta }{2}\right) \tanh \big(\frac{k _{\mbox{\tiny $0$}}}{2}\big) e^{\alpha_1 +(\k+2) x} \sin \om t +(\k+2) e^{2 (\alpha_1 +\k x)}+(\k-2)}\Bigg]\!,\quad\,\,\,\, \mbox{}
\er
where,
\br
\a_1 = \frac{1}{2}\ln\[\frac{ (\cosh k _{\mbox{\tiny $0$}}-\cos\zeta)(\cosh\eta +\cos\zeta)}{(\cosh k _{\mbox{\tiny $0$}}+\cos\zeta) (\cosh\eta-\cos\zeta)}\],\quad\mbox{}
\er
for $-\frac{\pi}{2}< \zeta <i\eta <\frac{\pi}{2}$. In the limit $\zeta\to i\eta$, we obtain the ground state configuration (\ref{eq6.2}), namely
\br
\vp_{\mbox{\tiny $B$}} \to 2\arctanh\left[e^{2 x} \tanh \left(\frac{\eta }{2}\right) \tanh \left(\frac{k _{\mbox{\tiny $0$}}}{2}\right)\].
\er


\subsection{Method of Images: Fermionic part}

Now, we study the fermionic part of the three-soliton solution. Let us first describe the situation with an ingoing and outgoing solitons where the fermionic parameters satisfy in particular that $\eps_1 = \eps_2 = \eps$ without loss of generality, and a stationary soliton at the origin described by the Grassmannian parameter $\eps_3$ being independent of the incoming and outgoing solitons. From the $\eps_3$-component of eq. (\ref{e4.51}) we obtain,
\br
 b_3 \left[\pa_t\(\frac{\mu_3 -\chi_3}{h(\vp_3)}\) - h(\vp_3)\(\mu_3+\chi_3\) \] =0.
\er
In the case of $b_3\neq 0$, the above relation give us the following constraints on the parameter,
\br
 k _{\mbox{\tiny $0$}} = \pm \eta  \quad \Longrightarrow  \quad a_3 =\pm 1 , \quad (a_1 a_2) =1 ,
\er
but it is inconsistent with the fermionic part of the stationary soliton solution, namely
\br
 \bpsi^{(3)} = 2 \eps_3 \(\frac{b_3}{a_3}\) \frac{(a_3e^{2x})}{1-(a_3e^{2x})^2}\Big|_{x=0, \,\, a_3 =\pm 1} \longrightarrow \infty.
\er
and also with the bosonic solution,
\br
 \vp_3(0,t) = 2\arctanh\left[\frac{
 \left(\frac{\cosh\th}{1-\cosh\th}\) F(t) -a_3(1+ \coth^2\(\th\o 2\))}{
\( 1 + \coth^2\(\th\o 2\)\) -  \left(\frac{\cosh\th}{1-\cosh\th}\)a_3\,F(t) }\]_{a_3=\pm 1} \longrightarrow\mp \,\infty.
\er
Then, a possible solution consistent with the bosonic boundary condition $\vp_3\big|_{x=0} = k _{\mbox{\tiny $0$}}$,  requires that $b_3=0$. It remains to determine the relation between the ingoing and outgoing soliton parameters $b_1$ and $b_2$. Then, the three-soliton solution take the following form,
\br
 \psi_3 = \eps \( \mu_1 \psi^{(1)} + \mu_2 \psi^{(2)}\),\label{eq6.41}\\
  \bpsi_3 = \eps \( \chi_1 \bpsi^{(1)} + \chi_2 \bpsi^{(2)}\).\label{eq6.42}
\er
where the functions $\mu_k$ and $\chi_k$ are given in the appendix \ref{appA}. Taking the $\eps$-projection of the eq. (\ref{e4.52}), we find the equation
\br
  \pa_t\(\frac{\psi_3 -\bpsi_3}{h(\vp_3)}\) - h(\vp_3)\(\psi_3+\bpsi_3\)  =0.
\er
Then after a long computation we find the following relation holds,
\br
 \(\frac{b_2}{a_2}\) = - e^{-\th} \( \frac{b_1}{a_1} \). \label{eq6.44}
\er
This relation reflects the symmetry used to derive the fermionic boundary potential, i.e. $\psi_2(x,t) \to -\psi_1(-x,t)$, and remembering that,
\br 
\bpsi^{(k)} = \s_k \psi^{(k)} \,=\, e^{\th_k}\, \psi^{(k)}.
\er

Now, let us consider a different situation where $\eps_1 \neq \eps_2$ but $\eps_3 = \nu_1 \eps_1 + \nu_2 \eps_2$, where $\{\nu_k\}_{k=1,2}$ are real arbitrary constants. Then, taking the $\eps_k$-projection of (\ref{e4.52}), namely{\small
\br
 \pa_t\[\bpsi^{(k)}\({\mu_k\o \s_k} - \chi_k\) + \bpsi^{(3)}\(c_k\mu_3- \chi_3\)\]\!\!\!&=&\!\!\! \left[\frac{H(\vp_3) (\pa_t\vp_3)(\mu_k-\s_k\chi_k) + 2h^2(\vp_3) (\mu_k +\s_k\chi_k)}{2\s_k}\]\!\bpsi^{(k)}\nonumber \\
 \mbox{}\!\!\! &+&\!\!\!\left[\frac{H(\vp_3) (\pa_t\vp_3)(c_k\mu_3-\chi_3) + 2h^2(\vp_3) (c_k\mu_3 +\chi_3)}{2}\]\!\bpsi^{(3)},\nonumber \\
\er}\normalsize
where $k=1,2$, we find that $\nu_1 =\nu_2 =1$ and the fermionic parameters of the stationary soliton satisfy the following relations,
\br
\(b_3^{+}\o a_3^{+}\) &=& \left[{\cosh\th +\cosh \eta}\o  (1+e^ \th) (1+\cosh \eta) \] \(b_1\o a_1\),  \quad \rm{or} \\
 \(b_3^{-}\o a_3^{-}\) &=& \left[{\cosh\th +\cosh k _{\mbox{\tiny $0$}}}\o  (1+e^ \th) (1+\cosh k _{\mbox{\tiny $0$}}) \] \(b_1\o a_1\),
\er
and
\br
 \(b_3^{+}\o a_3^{+}\) &=& \left[{e^\th\(\cosh\th +\cosh \eta\)}\o  (1+e^ \th) (1+\cosh \eta) \] \(b_2\o a_2\),  \quad \rm{or} \\
 \(b_3^{-}\o a_3^{-}\) &=& \left[{e^\th\(\cosh\th +\cosh k _{\mbox{\tiny $0$}}\)}\o  (1+e^ \th) (1+\cosh k _{\mbox{\tiny $0$}}) \] \(b_2\o a_2\),
\er
which imply that the following relation between the amplitudes of the ingoing and outgoing soliton parameters is always satisfied,
\br
  \(\frac{b_2}{a_2}\) =  e^{-\th} \( \frac{b_1}{a_1} \).
\er

Finally, it is worth examining what form the fermionic part of the solution takes when we are considering boundary breathers on the half-line. To do that in the present setting, it is natural to set $\eps_1=\eps_2$, and considering the choice of parameters in (\ref{beq 6.32}) for the bosonic part and the relation (\ref{eq6.44}), we find that $b_2=e^{-i\zeta}\,b_1$. Now, the ``fermionic boundary breathers'' take the following form,
\br 
\psi_3 &=&\eps_1 b_1\!\left[{u_0(x) + u_{1}(x)\cos(\om t) + u_{-1}(x)\sin(\om t)+ u_{2}(x)\cos(2\om t)   \o v_0(x)+ v_1(x)\sin(\om t) + v_2(x)\cos(2\om t)} \right],\\[0.2cm]
\bpsi_3 &=& -\eps_1 b_1\!\left[{{u}_0(x) - {u}_{1}(x)\cos(\om t) + {u}_{-1}(x)\sin(\om t) + {u}_{2}(x)\cos(2\om t)   \o {v}_0(x)+ {v}_1(x)\sin(\om t) + {v}_2(x)\cos(2\om t)} \right],
\er
where the $x$-dependent coefficients are listed down,
\br
{u}_{0}(x) &=& -96\,i \,e^{(\a_1+\b)}\,e^{2i\z}\cos^2\z\, e^{2(\k x+1)},\\[0.2cm]
u_{1}(x) &=& -4ie^{\k x} \,e^{i\z}\sin(2\z)\left[e^{4x}\big((1+e^{i \zeta })^2+  (1- e^{i \zeta } )^2 e^{2 (\kappa x+ \alpha _1)} \big)\nonumber \right. \\ && \left.   \qquad \qquad \qquad \qquad  -\,e^{2\b}\big((1-e^{i \zeta })^2+(1+e^{i \zeta })^2 e^{ 2(\kappa x+ \alpha _1)} \big) \right],\\[0.2cm]
u_{-1}(x) &=& 4i\,e^{\k x}\cos\z\left[e^{4x}\big((1+e^{i \zeta })^4 -  (1-e^{i \zeta })^4 e^{2 (\kappa x+ \alpha _1)} \big) \right.\nonumber \\ && \left.\qquad \qquad \qquad \quad  - e^{2\b}\big( (1-e^{i \zeta })^4-  (1+e^{i \zeta })^4 e^{ 2(\kappa x+ \alpha _1) } \big)\right],
\er
\br
u_2(x) &=& 32\,i\,e^{\a_1+\b}\,e^{2i\z}\cos^2\z\,e^{2(\k+1)x},\\[0.2cm]
v_0(x) &=&  (e^{i \zeta }-1)\left[e^{2 \beta } \big((e^{i \zeta }-1)^4+e^{4 \left( \kappa x+\alpha _1\right)} (1+e^{i \zeta })^4-8 e^{2i\z}\,e^{2(\kappa x +\alpha _1)}\big)\right. \nonumber \\ && \left. - e^{4x}\big((1+e^{i \zeta })^4 + e^{4 \left(x \kappa +\alpha _1\right)} (e^{i \zeta }-1)^4- 8 e^{2i\z}\,e^{2(\kappa x+\alpha _1)}\big)\right],\\[0.2cm]
v_1(x) &=& -64 \,e^{\b}\,e^{2i\z}(e^{i\z}-1)\cos\z \,e^{2((\k+1)x+\a_1)}\sinh(\k x+\a_1), \\[0.2cm]
v_2(x) &=& -8 \,e^{2i\z}(e^{i\z}-1)\cos^2\z\,e^{2(\k x +\a_1)}(e^{4x}-e^{2\b}).
\er


\section{Conclusions}

In this paper, we have investigated the necessary conditions satisfied by the boundary parameters to preserve both integrability and supersymmetry of ${\cal N}=1$ SShG model in the half line, by applying a limiting procedure to its corresponding defect theory. 

The three-soliton solution for ${\cal N}=1$ SShG model has been constructed by employing the B\"acklund transformation, and then used to perform a classical analysis in the half line. Explicit expressions for boundary breathers solutions have been presented as well as the relations satisfied by the respective boundary parameters.
 
It would be interesting to derive the three-soliton solutions for ${\cal N}=2$ SShG model by extending the procedure proposed in \cite{Lean2010}, in order to investigate the model on a half line. This problem will be addressed in future developments.


\vskip 1cm
\noindent
{\bf Acknowledgements} \\
\vskip .1cm \noindent
{ARA would like to thank to  FAPESP S\~ao Paulo Research Foundation
for financial support under the PD fellowship 2012/13866-3. JFG and AHZ thank CNPq for financial
support.}


\appendix
\section{Three-soliton solution}
\label{appA}

The three-soliton solution $\P_3\equiv \P^{(1,2,3)}$ for the ${\cal N} =1$ SShG can be constructed as indicated in the Bianchi diagram,
\begin{eqnarray}
\begin{diagram}\dgARROWLENGTH=2.2em
 \node{} \arrow[2]{s,!}\node[4]{\P^{(1)}}\arrow[2]{ese,t,1}{\s_2, \,\,F_{(1)}^{(1,2)} }\\[2]
 \node{\P_0}\arrow[2]{ene,t}{\s_1,\,\,F_{(0)}^{(1)} } \arrow[2]{ese,t,2}{\s_2,\,\,F_{(0)}^{(2)} } \node[8]{\P^{(1,2)}}\arrow[2]{ese,t}{\s_3,\,\,F_{(1,2)}^{(1,2,3)}}\\[2]
 \node{}\node[4]{\P^{(2)}} \arrow[2]{ene,b}{\s_1,\,\,F_{(2)}^{(1,2)} } \arrow[2]{ese,b}{\s_3,\,\,F_{(2)}^{(2,3)}} \node[8]{\P^{(1,2,3)}}\\[2]
 \node{\P_0}\arrow[2]{ene,t}{\s_2,\,\,F_{(0)}^{(2)} } \arrow[2]{ese,b}{\s_3,\,\,F_{(0)}^{(3)} } \node[8]{\P^{(2,3)}}\arrow[2]{ene,b}{\s_1,\,\,F_{(2,3)}^{(1,2,3)}}\\[2]
 \node{} \node[4]{\P^{(3)}} \arrow[2]{ene,b}{\s_2,\,\,F_{(3)}^{(2,3)}}
\end{diagram}\label{Fig A.1}
\end{eqnarray}
by using appropriately the B\"acklund transformations,
\begin{eqnarray}
  D_{\bar z}(\P_0-\P) &=& -\sqrt{2\s}\, F\cosh\( \frac{\P_0+\P}{2}\),\label{2.1}\\
  D_z(\P_0+\P) &=& i\sqrt{\frac{2}{\s}} \,F\cosh\( \frac{\P_0-\P}{2}\),\label{2.2}\\
  D_{\bar z} F &=&\sqrt{2\s}\sinh\(\frac{\P_0+\P}{2}\),\\
  D_z F &=&i\sqrt{\frac{2}{\s}} \sinh\(\frac{\P_0-\P}{2}\),\label{2.3}
\end{eqnarray}
where $F$ is an auxiliary fermionic superfield,  $\s$ the B\"acklund parameter, and the superderivatives given by
\br
\quad D_{\bar z}&=&\frac{\pa}{\pa \th_1}+\th_1\pa_{\bar z}, \qquad \mbox{and}  \qquad  D_z \,=\, \frac{\pa}{\pa \th_2}+\th_2\pa_z.
\er
Here, we have used the light-cone coordinates $z=x-t$, ${\bar z}=x+t$, $\pa_{\bar z} =  \frac{1}{2}\(\pa_x+\pa_t\)$ and $\pa_z = \frac{1}{2}\(\pa_x-\pa_t\)$. Then, the equations of motion  (\ref{mov1}) can be written in the superfield notation as,
\br 
 D_{\bar z}D_z \P= i\sinh\P .
\er
So starting from the vacuum configuration $\P_0=0$, the three-soliton solution $\P_3$ can be written in components is the following form,
\br
\p_3 &=& \p^{(2)} + 2\arctanh\left[\d_{13}\tanh\t_3 \right] - \rho_{12}\bpsi^{(1)}\bpsi^{(2)} +\rho_{13}\bpsi^{(1)}\bpsi^{(3)} -\rho_{23}\bpsi^{(2)}\bpsi^{(3)},\\[0.2cm]
\bpsi_3 &=& \chi_1 \bpsi^{(1)} +  \chi_2 \bpsi^{(2)} + \chi_3\bpsi^{(3)} + \chi_4 (\bpsi^{(1)}\bpsi^{(2)}\bpsi^{(3)}),\\[0.2cm]
\psi_3 &=& \mu_1\psi^{(1)} +  \mu_2 \psi^{(2)} + \mu_3\psi^{(3)} + \mu_4 (\psi^{(1)}\psi^{(2)}\psi^{(3)}),
\er
where we have defined $\t_3= (\vp^{(1,2)}-\vp^{(2,3)})/2$, and by introducing $\l_1^{\pm}=(\p^{(2)}\pm\vp^{(1,2)})/2$ and $\l_2^{\pm}=(\p^{(2)}\pm\vp^{(2,3)})/2$, we get {\small
\br
\rho_{12}&=& \left[\frac{\xi^{(1,2)}(\xi^{(2,3)}-1)\,\sech\,\l_1^{+}\, \sech\,\l_2^{+}\,}{4\sqrt{\sigma_1 \sigma_3}}\right]\left[\frac{A_{13}\sinh\t_3}{B_{13}-\sinh^2\t_3}\right] +\left[\frac{\delta_{13}\,\sech^2 \t_3\,  f^{(1,2)} }{1-\delta_{13}^2\tanh^2 \t_3 } \right] \\[0.5cm]
\rho_{13} &=& -\left[\frac{\xi^{(1,2)}\xi^{(3,2)}\,\sech\,\l_1^{+}\, \sech\,\l_2^{+}\, }{4\sqrt{\sigma_1 \sigma_3}} \] \left[\frac{A_{13}\sinh\t_3 }{B_{13}-\sinh^2\t_3 }\right]\\[0.5cm]
\rho_{23} &=& \left[\frac{\xi^{(3,2)}(\xi^{(2,1)}-1) \,\sech\,\l_1^{+}\, \sech\,\l_2^{+}\,}{4\sqrt{\sigma_1 \sigma_3}}\right]\left[\frac{A_{13}\sinh\t_3}{B_{13}-\sinh^2\t_3}\right]-\left[\frac{ \delta_{13}\,\sech^2\t_3\,f^{(2,3)}}{1-\delta_{13}^2\tanh^2\t_3}\right],  \qquad\mbox{}
\er
\br
\chi_1 &=& \left\{\left[\sqrt{\frac{\sigma_3}{\sigma_1}}\,\frac{\sinh\l_2^{+} \,\sech\,\l_1^{+}}{2}\right]\left[\frac{A_{13}\sinh\t_3}{B_{13}-\sinh^2\t_3}\right]+\frac{ \delta_{13}\,\mathrm{sech}^2\t_3}{1-\delta_{13}^2\tanh^2\t_3}\right\}\xi^{(1,2)}, \qquad\mbox{}\\[0.5cm]
\chi_2 &=& 1 - \frac{ \delta_{13}\,\mathrm{sech}^2\t_3}{1-\delta_{13}^2\tanh^2\t_3}\(\xi^{(2,3)} -\xi^{(2,1)}\)-\frac{1}{2}\sqrt{\frac{\s_3}{\s_1}}\left[\frac{A_{13}\sinh\t_3}{B_{13}-\sinh^2\t_3}\right]\,\sinh\l_2^{+} \, \sech\,\l_1^{+}\, \big( 1-\xi^{(2,1)}\big)\nonu \\
 &&+\frac{1}{2}\sqrt{\frac{\s_1}{\s_3}}\left[\frac{A_{13}\sinh\t_3}{B_{13}-\sinh^2\t_3}\right]\, \sinh\l_1^{+} \,\sech\,\l_2^{+}\,\big(1-\xi^{(2,3)} \big) \\[0.4cm]
\chi_3 &=&-\left\{ \left[\sqrt{\frac{\s_1}{\s_3}}\frac{\sinh\l_1^{+} \, \sech\,\l_2^{+}\,}{2}\right]\left[\frac{A_{13}\sinh\t_3}{B_{13}-\sinh^2\t_3}\right]+\frac{ \delta_{13}\,\mathrm{sech}^2\t_3}{1-\delta_{13}^2\tanh^2\t_3}\right\}\xi^{(3,2)}\qquad \mbox{} \\[0.4cm]
\chi_4 &=& \left[\frac{1}{4}\sqrt{\frac{\s_1}{\s_3}} \frac{A_{13}\sech\,\l_2^{+} }{B_{13}-\sinh^2\t_3}
\right]\left[  \frac{B_{13}\sinh(\l_1^{+}+\t_3)+\sinh^2\t_3\sinh(\l_1^{+}-\t_3) }{\left(B_{13}-\sinh^2\t_3 \right)}\right] \(\xi^{(3,2)} f^{(1,2)}\) \nonu \\
&&+ \left[\frac{1}{4}\sqrt{\frac{\s_3}{\s_1}} \frac{A_{13}\,\sech\,\l_1^{+}}{B_{13}-\sinh^2\t_3} \]
\left[  \frac{B_{13}\sinh(\l_2^{+}-\t_3)+\sinh^2\t_3\sinh(\l_2^{+}+\t_3)}{\left(B_{13}-\sinh^2\t_3\right)}   \right]  \( \xi^{(1,2)} f^{(2,3)}\)\nonu\\
 &&-\left[\frac{ \delta_{13}\big(1-\d_{13}^2\big)\mathrm{tanh}\t_3\,\mathrm{sech}^2\t_3}{\left(1-\delta_{13}^2\tanh^2\t_3\right)^2} \right] \Big( \xi^{(1,2)} f^{(2,3)} +\xi^{(3,2)} f^{(1,2)} \Big)   \\[0.4cm]
 \m_1 &=& \left\{\left[\sqrt{\frac{\s_1}{\s_3}}\frac{\sinh\l_2^{-} \,\sech\,\l_1^{-} }{2}\]\left[\frac{A_{13}\sinh\t_3}{B_{13}-\sinh^2\t_3}\right]+\frac{ \delta_{13}\,\mathrm{sech}^2\t_3}{1-\delta_{13}^2\tanh^2\t_3}\right\} \eta^{(1,2)}, \qquad\mbox{} \\[0.4cm]
\m_2 &=& 1 - \frac{ \delta_{13}\mathrm{sech}^2\t_3}{1-\delta_{13}^2\tanh^2\t_3}\big(\eta^{(2,3)}-\eta^{(2,1)}\big)-\left[\sqrt{\frac{\s_3}{\s_1}}\frac{\sinh\l_1^{-} \sech\,\l_2^-}{2}\]\left[\frac{A_{13}\sinh\t_3}{B_{13}-\sinh^2\t_3}\right] \big(1+\eta^{(2,3)} \big) \nonu \\
 &&+\left[\sqrt{\frac{\s_1}{\s_3}}\frac{\sinh\l_2^-  \sech \l_1^-}{2}\]\left[\frac{A_{13}\sinh\t_3}{B_{13}-\sinh^2\t_3}\right] \big(1-\eta_2^{(1,2)} \big)\\[0.4cm]
\m_3 &=&-\left\{\left[\sqrt{\frac{\s_3}{\s_1}} \frac{\sinh\l_1^{-} \sech\,\l_2^-}{2}\]\left[\frac{A_{13}\sinh\t_3}{B_{13}-\sinh^2\t_3}\right] +\frac{ \delta_{13}\,\mathrm{sech}^2\t_3}{1-\delta_{13}^2\tanh^2\t_3} \right\}\eta^{(3,2)},\qquad\mbox{} \\[0.4cm]
\m_4 &=&  \left[\frac{1}{4}\sqrt{\frac{\s_3}{\s_1}} \frac{A_{13} \sech\,\l_2^- }{B_{13}-\sinh^2\t_3} \] \left[ \frac{ B_{13}\sinh( \l_1^- -\t_3) +\sinh^2\t_3 \sinh( \l_1^- +\t_3 ) }{\left( B_{13}-\sinh^2\t_3 \right)}   \right] \big( \eta^{(3,2)} f^{(1,2)} \big) \nonu\\
&&- \left[\frac{1}{4}\sqrt{\frac{\s_1}{\s_3}}\frac{A_{13}\,\sech\,\l_1^-}{B_{13}-\sinh^2\t_3}\]\left[ \frac{B_{13}\sinh(\l_2^-+\t_3)+\sinh^2\t_3 \sinh(\l_2^- -\t_3) }{\left(B_{13}-\sinh^2\t_3 \right)} \right] \big( \eta^{(1,2)} f^{(2,3)}\big)\nonu\\
&&-\left[ \frac{ \delta_{13}\big(1-\d_{13}^2\big) \mathrm{tanh}\t_3\,\mathrm{sech}^2\t_3}{\left(1-\delta_{13}^2\tanh^2\t_3 \right)^2}  \right] \big(\eta^{(1,2)} f^{(2,3)}+\eta^{(3,2)} f^{(1,2)} \big). 
\er
}



\end{document}